# Experimental observation of strong photon localization in disordered photonic crystal waveguides

J. Topolancik[1], B. Ilic[2], and F.Vollmer[1]

[1]Rowland Institute at Harvard, Harvard University, Cambridge, MA 02142

[2]Cornell Nanofabrication Facility and Field of Biophysics, Cornell University, Ithaca, NY 14853

## Abstract

We demonstrate experimentally that structural perturbations imposed on highly-dispersive photonic crystal-based waveguides give rise to spectral features that bear signatures of Anderson localization. Sharp resonances with the effective Qs of over 30,000 are found in scattering spectra of disordered waveguides. The resonances are observed in a ~20-nm bandwidth centered at the cutoff of slowly-guided Bloch-modes. Their origin can be explained with interference of coherently scattered electromagnetic waves which results in the formation of a narrow impurity (or localization) band populated with spectrally distinct quasistates. Standard photon localization criteria are fulfilled in the localization band.

Transport of waves through a medium can be severely suppressed and even halted by interference and multiple-scattering from random impurities which can give rise to strong (or Anderson) localization[1-3]. The theory behind the process was originally developed for matter-waves (electrons in disordered atomic crystals)[1], but it can be directly extended to classical waves where the verification of its universal principles is not complicated by coulomb and electron-phonon interactions[2,3]. In this regard, random scattering media[4-7] and disordered lattices[8] have attracted considerable experimental interest as promising model systems for testing the localization concepts on electromagnetic waves and light in particular. Strong localization of light, though rather difficult to achieve, has been successfully demonstrated in highly-scattering random media[6,7] and, more recently, in a disordered two-dimensional (2D) lattice[8]. It has been proposed by John[9] that the localization conditions should be easier to fulfill in high-index-contrast periodic structures with a photonic bandgap (PBG)[10]. To our knowledge this idea has been tested successfully only in metal-wire networks at microwave frequencies where absorbing materials complicate interpretation of results as it is difficult to distinguish absorption from localization[11]. In this letter, we report direct experimental observations of localized electromagnetic fields in non-absorbing, geometrically-disordered photonic crystal (PhC) structures at optical frequencies.

High-refractive-index slabs with 2D arrays of air-holes exhibit large PBGs for the TE-like polarization (**E**-field parallel to 2D-plane)[12], which established them as a popular platform for designing waveguides[13-15] and nanocavities[16-18]. Engineered PhC nanostructures guide and confine light remarkably well by relying on Bragg reflections in the 2D-plane and on total internal reflection in the out-of-plane direction. Guiding losses and cavity quality (Q) factors are believed to be limited primarily by the fabrication-induced surface roughness which breaks the PhC periodicity and destroys the Bloch-wave coherence by diffusive scattering. Considerable efforts have thus been expended to improve fabrication processes which, together with the progress in design optimization, have produced highly-dispersive (or slow-light) waveguides[19,20] and optical nanocavities with ultra-high Qs and record-low modal volumes[21]. Here we show experimental evidence that random departures from index-periodicity need not necessarily set off decoherence, but can produce a fundamentally different type of coherent localization analogous to that observed in strongly-scattering random media[4-7]. More specifically, we will argue that scattered spectra collected from geometrically-disordered PhC waveguides in free-standing silicon slabs bear signatures of Anderson localization.

Disordered PhC structures consisting of a hexagonal lattice of holes were patterned in silicon-on-insulator substrates using electron beam lithography and chlorine-based inductively coupled plasma reactive ion etching. PhCs with various fill factors ($f = \langle r \rangle / a$, where $\langle r \rangle$ is the effective hole radius and $a$ is the lattice constant) were fabricated. As illustrated in the scanning electron micrograph (SEM) in Fig. 1(b), the fabricated patterns carry a significant geometrical disorder in addition to the usual surface roughness introduced during fabrication. The air-holes form an array with a lattice constant $a$=410 nm; their size ($Area = \pi \langle r \rangle^2$) is fairly constant with a standard deviation of 3.6%; but their shape deviates noticeably from an ideal circle. More explicitly, the roundedness (or circularity) of the holes, defined as $4\pi \times Area / Perimeter^2$, is 0.85 whereas values very close to unity are readily achieved with state-of-the-art lithographic tools. It will be argued later that the introduced geometrical perturbations are small enough not to significantly affect the band structure of the underlying periodic lattice (this way the traditional PBG defect engineering concepts such as bandgap, point-defect mode, line-defect dispersion, etc. still apply), but sufficient to generate significant multiple-scattering of Bloch-waves necessary for strong localization. Once the holes were etched into the ~210-nm-thick silicon layer, the patterns were cleaved and the buried oxide layer was removed with a buffered hydrofluoric acid solution forming a free-standing PhC slab shown in Fig. 1(c). The inner walls of the etched holes are smooth and nearly vertical. The disordered PhC enfolds a ~60-μm-long line-defect waveguide (W1) formed by a row of missing holes along the $\Gamma K$ direction of the reciprocal-lattice, surrounded on both sides by ten rows of holes (see Figs. 1(b) and (c)). Various donor-defect cavities defined by a single or multiple missing holes were also patterned near the waveguides in some samples. An example of a linear, three-defect cavity is shown in Fig. 1(e).

Coherent light from an infrared (IR) diode laser tunable from 1,475 to 1,580 nm, was coupled into W1s from a single-mode optical fiber (SMF-28). To compensate for the significant impedance mismatch inherent to conventional end-fire coupling, PhC modes were excited with a non-linear fiber taper. The taper, prepared by pulling a melted fiber and etching its tip down to the W1 dimensions ($\sqrt{3} \times a$), was positioned on top of the PhC-slab as illustrated in Fig. 1(a). The arrangement allows the light to leak out of the taper and to evanescently couple into W1. Once excited, the PhC modes propagate in the waveguide and interact with cavities which leak the light vertically out of the slab. This light was collected with an infinity-corrected objective (100×, NA=0.80) and its intensity monitored with an InGaAs photodiode as the coherent source was scanned. A beam-splitter redirected a fraction of the collimated beam to an IR camera for imaging. A field stop was placed in front of the photodiode to locally probe ~10μm-long waveguide sections and to block parts of the free-propagating beam deflected accidentally into the objective from surface impurities.

The spectrum collected from a donor-defect cavity separated from W1 by four rows of holes is shown in Fig. 2(a). It contains two distinct features: an isolated peak at 1,492 nm with the Q of ~3,000 (Fig. 2(b)), as inferred from its full width at half maximum linewidth; and a striking series of extremely sharp ($\Delta\lambda$~50 pm), discrete peaks in a narrow band centered at ~1,570 nm (Fig. 2(d)). The IR image in Fig. 2(c) indicates that the analyzed light is emitted primarily from the donor-cavity region. To explain the origins of the measured spectra we need to address the dispersion characteristics of PhC waveguides and describe how they are affected by disorder. Although it is possible to systematically study how specific types of disorder affect transport through waveguides, e.g. by analyzing ensembles of randomized structures with the finite-difference-time-domain (FDTD) method[22], this would require unreasonably large simulation domains and unfeasible computation times. Such a methodical treatment of disorder is beyond the scope of this experimental paper and will be discussed elsewhere. Instead, here we use the established supercell approach and 3D plane-wave expansion[23] to calculate the band structure of line-defects in ideal, disorder-free crystals; and qualitatively explain how disorder affects dispersion and gives rise to the observed spectral features. The supercell used to compute the band diagrams is shown in the inset of Fig. 3(a). Its dimensions are $7\sqrt{3}a \times 4a \times a$ and the refractive index of silicon[24] used in the simulations is $n$=3.52. The simulated slab is $\sim 0.51 \times a$-thick and the holes are circular with radii that correspond to the fill factors of the fabricated PhCs determined from SEM images. The Figure 3(a) shows the projected band-structure of a W1 with $f\cong 0.28$. To conceptually show the effect of disorder we introduce error bars representing uncertainty in the computed eigenfrequencies. Their magnitude in Fig. 3(a) is arbitrary and is merely meant to reflect the severity of disorder, i.e. how much the departures from holes' circularity affect the band structure uncertainty. A single non-leaky mode bounded by the light-line ($\omega \cong 0.283[a/\lambda]$) and the stop-band ($\omega \cong 0.265[a/\lambda]$) falls within the scan-range of the probing laser. The mode has an even parity and exhibits anomalous dispersion unique to PhC waveguides[13-15]. Its group velocity ($v_g = d\omega/dk$) gradually decreases as the wave-vector approaches the zone boundary (the slow-light regime). We attribute the solitary spectral peak at 1,492 nm ($\omega \cong 0.2748[a/\lambda]$) to a point-defect cavity mode excited evanescently by the waveguide in the classical (or index-guided) regime. Even though the crystal disorder reduces the PBG and degrades the cavity Q, it permits proper waveguiding and Bragg localization. Whereas the isolated resonance can be accounted for with the conventional donor-cavity PBG defect picture[10], we do not believe the sharp peaks at longer wavelengths (in the slow-light regime) can be explained within the simple framework. Instead, we contend here that these features are caused by disorder and are a manifestation of Anderson localization.

Calculated dispersion diagrams of even modes in W1 waveguides with different fill factors: $f$=0.26, $f$=0.28 and $f$=0.30 are presented in Fig. 3(b). The plots indicate that increasing $f$ shifts the mode's edge to higher frequencies. The same trend is observed experimentally for the bands of narrow peaks measured on the fabricated disordered waveguides with equivalent $f$s (Fig. 3(c)), which suggests that the spectral position of the feature is dictated by the band structure. Our measurements directly probe the spectral characteristics of local fields in the PhC structure by analyzing the light scattered vertically by donor-cavities. The cavities are usually off-resonance in the bandwidth where the peaks are present and they merely enhance the scattering efficiency out of the PhC slab. Distinct Bragg defect-modes, such as the one shown in Fig. 2(b), are seldom detected as their spectral positions and Qs depend sensitively on the cavity-geometry and disorder generally either degrades the Q or shifts the resonant frequency outside the limited scan-range of the probing laser. On the other hand, pronounced narrow spectral features always appear, even in disordered waveguides fabricated without donor-cavities. In this case the sharp peaks are observed, though with a reduced intensity, in arbitrary locations along the probed waveguides. A close-up of the spectrum collected from a section of a disordered W1 with no donor-cavity ($f\cong0.30$) is presented in Fig. 4. The scan reveals a band with multiple sharp resonances, rather than a broad spectral feature anticipated as a result of the increased vertical-scattering loss from the waveguide in the slow-light regime[25]. The highly-coherent character of the detected light suggests that the introduced geometrical disorder changes the nature of scattering in this narrow band: diffusive scattering due to surface roughness is suppressed, being dominated by coherent scattering that leads to localization. As a result, a localization window, outlined as a dark-shaded area around the mode's edge in Fig. 3(a), opens in the $k-\omega$ space. The physical origins of strong localization in disordered W1s can be explained within the context of theories of wave propagation in disordered media[26-28].

Slow light in the highly-dispersive waveguides is susceptible to coherent backscattering which can block the transport if certain conditions are met. It has been proposed by John[9] that the standard Ioffe-Regel criterion for localization ($k\ell \leq 1$)[26] should be reinterpreted for periodic-index media in the following way: $k$ is the wave vector of a coherent Bloch wave and $\ell$ is its scattering mean free path. The magnitude of $k$ is limited by PhC periodicity attaining the maximum at the zone boundary $\left(|k_{max}| = \pi/a\right)$. $\ell$ represents the length-scale at which the mode is scattered by structural imperfections of the lattice and must therefore depend on dispersion. In the classical regime (large $v_g$) $\ell$ is large since the light is well confined within the waveguide and interacts only weakly with disorder. However, in the slow-light regime $\left(v_g \to 0\right)$ the guided mode probes the crystal with progressively increasing evanescent fields, which enhances scattering from disorder and reduces $\ell$ as $k$ approaches the zone boundary. The strong-localization window opens when $v_g$ (and therefore $\ell$) becomes sufficiently small to satisfy the modified Ioffe-Regel criterion.

Alternatively, it can be argued that introduction of random disorder fills the edge of the stop-band with quasistates creating a string of resonant cavities along the waveguide. The defect states that populate the stop-band are well-localized, i.e. spatially and spectrally distinct, only if their level spacing $(\Delta \nu)$ is large enough and the level widths $(\delta \nu)$ small enough so that the modes do not overlap. This essentially says that another fundamental localization condition, the Thouless criterion ($\delta \equiv \delta\nu/\Delta\nu < 1$)[27], is satisfied. Significantly-overlapping modes would enable transport and destroy localization. The origin of the localization band is shown schematically in the inset of Fig. 4. A disorder-free W1 exhibits an abrupt transition from the guided mode to the stop-band, i.e. the density-of-states (DOS) of the guided modes, $\rho(\omega) \propto \left(v_g\right)^{-1}$, diverges at the mode's cutoff beyond which it suddenly vanishes. Disorder causes band-structure fluctuations that smear the sharp cutoff creating a transitional (or impurity) band filled with both, slowly-guided modes credited to the residual refractive-index periodicity and localized quasistates arising from disorder. Light-propagation in the band can be viewed as a combination of remnant waveguiding and resonant transport. The observed localized quasimodes with effective Qs of over 30,000 (Fig. 4) are in many respects similar to the engineered defect-modes in PhC-heterostructure cavities[21] in which periodicity of PhC waveguides is broken intentionally by locally increasing the lattice constant. These modes have small modal volumes and record-high Qs of up to $\sim 10^6$. An interesting question worth exploring is whether random disorder can confine light with similar efficiency.

Disordered PhC waveguides in high-index-slabs are ideal for systematic studies of Anderson localization. Slow light is coherently localized in these structures as a result of "a delicate interplay between order and disorder" envisioned by John[9]. The nature and the level of disorder, and hence the onset of localization, can be controlled lithographically, while the spatial and spectral characteristics of the localized fields can be obtained with near-field optical probes. Critical localization parameters such as the localization volume, the Thouless number, conductance etc. can thus be precisely determined with direct measurements. Disordered PhCs might also find various technological applications as low-threshold lasers, ultra-sensitive detectors and unique barcode labels.

**Acknowledgements**

This work was supported by the Rowland Junior Fellowship program and was performed in part at the Center for Nanoscale Systems (CNS) which is supported by the National Science Foundation under NSF award no. ECS-0335765.

**Figure Legends**

**Figure 1**

**a)** Schematic of the measurement setup. Vertically scattered light is collected with an objective (O) and imaged with a lens (L1) onto a field-stop (FS) consisting of a variable aperture which selects the field of view. Another lens (L2) focuses light from the selected area into an IR photodiode (PD). A beam splitter redirects a fraction of the collimated beam into an IR charge coupled device (IR-CCD) for imaging. **b)** SEM of a typical disordered PhC pattern; **c)** the cross-section of a free-standing W1 PhC waveguide**; d)** top image of a W1 waveguide; and **d)** example of a donor-defect cavity.

**Figure 2**

**a)** Spectrum acquired from a single-defect cavity separated from the W1 waveguide by four rows of air-holes; **b)** 4nm-wide detailed scan showing the solitary broad feature; **c)** IR image of the collected light; and **d)** high-resolution scan of a 4nm-wide section of the band containing sharp spectral peaks.

**Figure 3**

**a)** Calculated band structure of a W1 waveguide ($f \cong 0.28$) with an even-parity guided-mode (black circles). The dark-shaded region around the mode's edge outlines the strong-localization window. The inset shows the supercell used in the plane-wave expansion simulations. **b)** Calculated dispersion curves for the even mode in waveguides with three different fill factors; and **c)** spectra collected from fabricated waveguides with the equivalent fill factors. The insets show the corresponding SEM images of the PhC structures.

**Figure 4**

Detailed spectrum of the localization band collected from a disordered W1 ($f \cong 0.30$). The inset shows the calculated DOS of a defect-free waveguide. Disorder creates a localization band (gray) around the slow-mode's cutoff.

**Figure1**

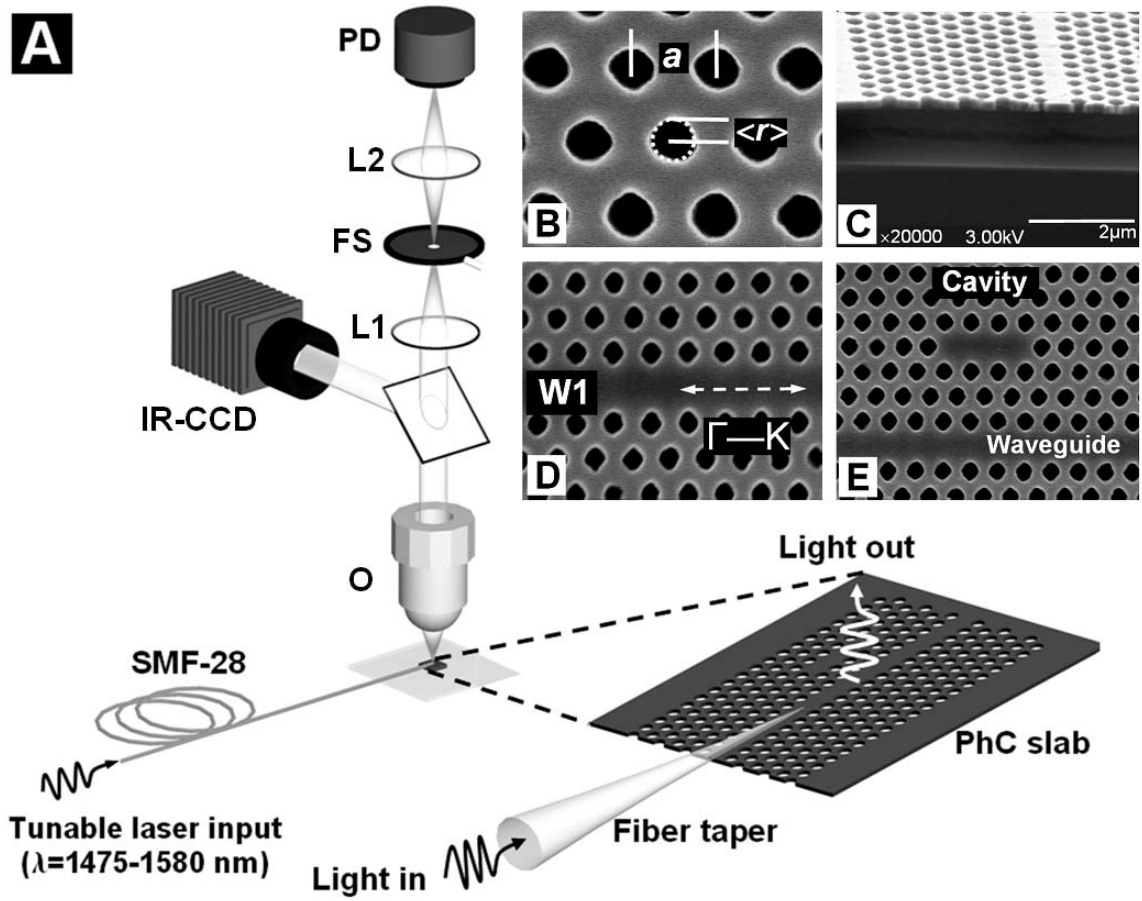

**Figure 2**

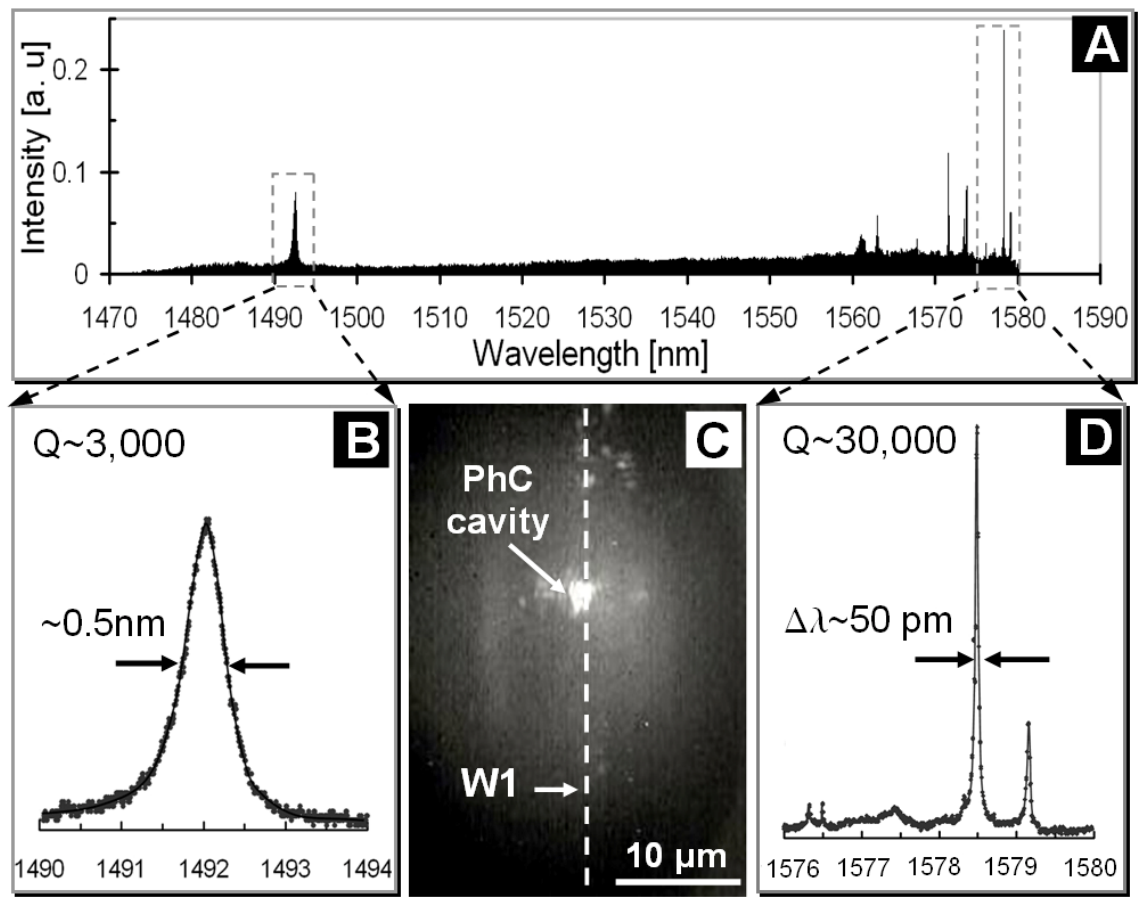

**Figure 3**

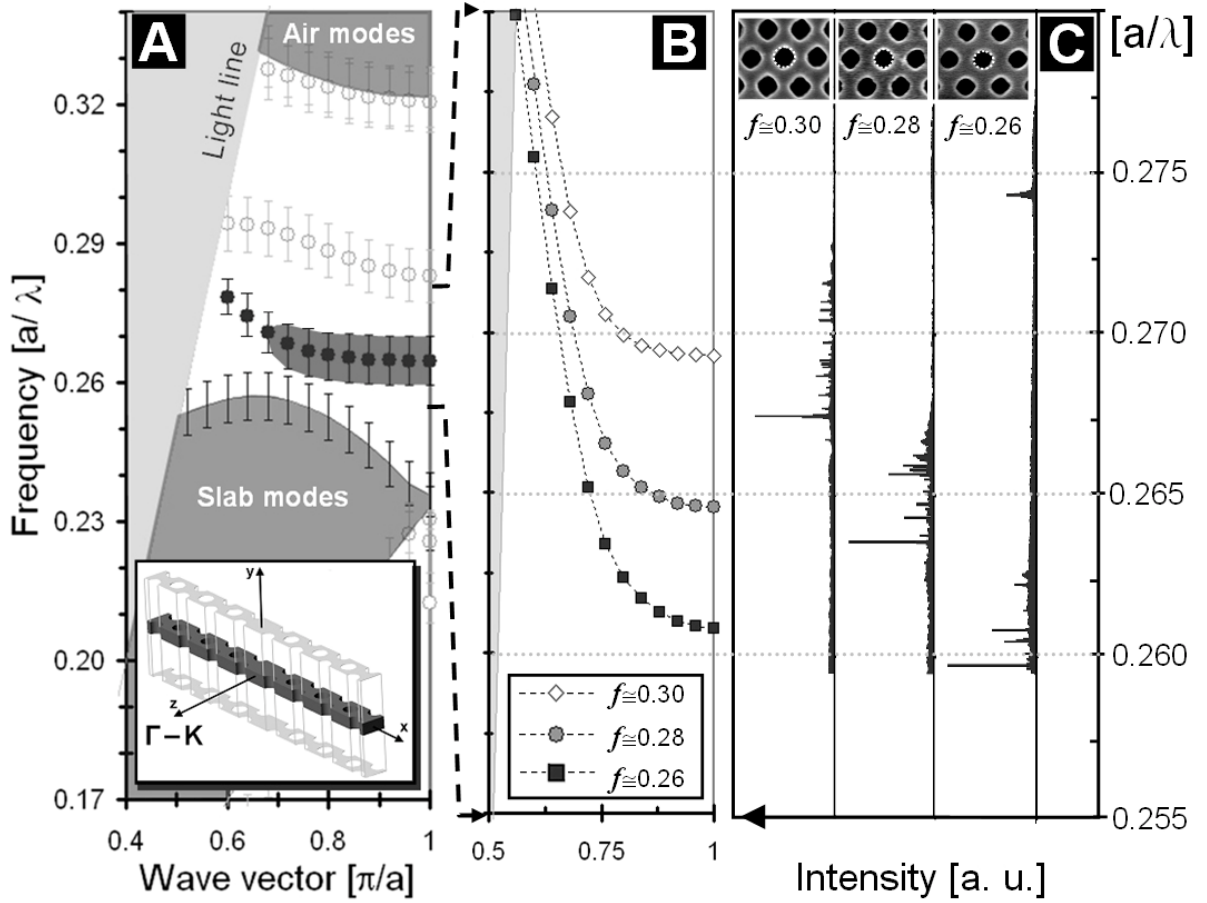

**Figure 4**

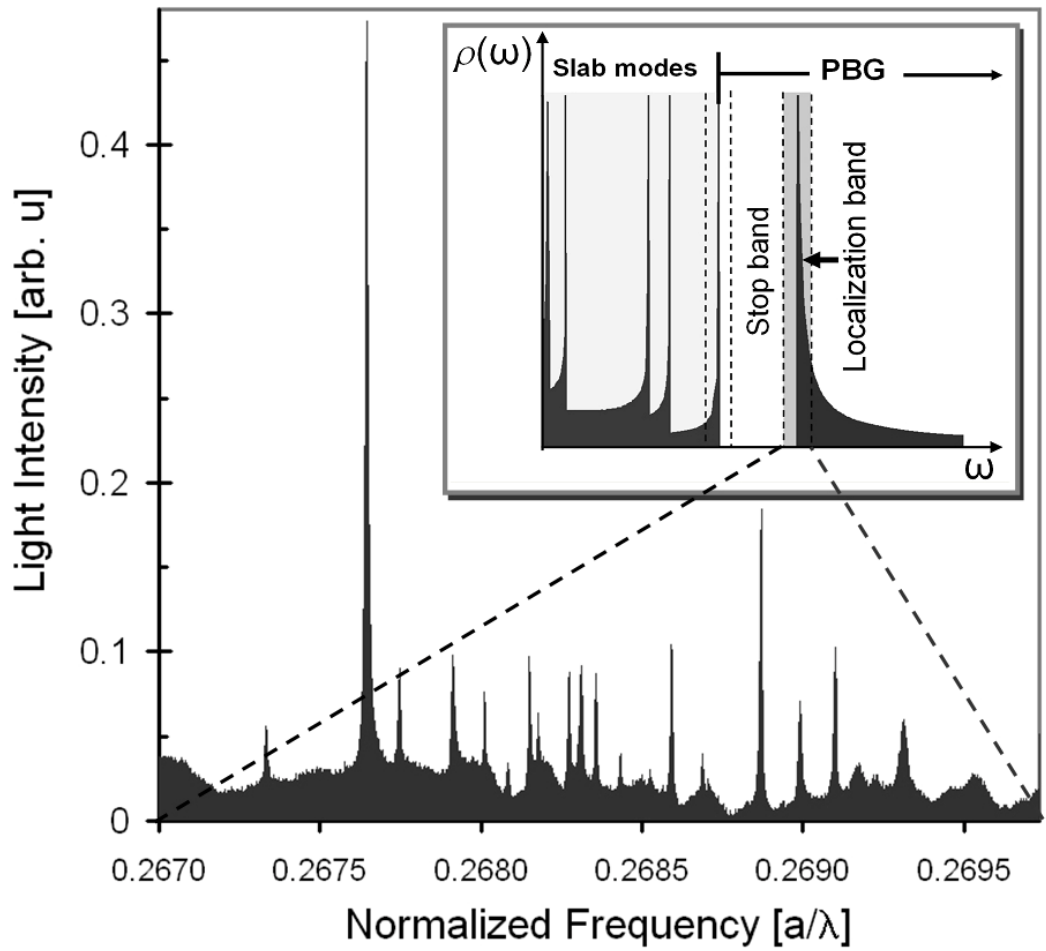